\documentclass[preprint,12pt]{elsarticle}
\usepackage{float}

\usepackage{caption}
\usepackage{physics,amsmath}
\usepackage{subcaption}
\usepackage{graphicx}
\usepackage{multirow}
\usepackage{layouts}
\usepackage{pifont}
\usepackage[left=2cm, right =2cm, top=3.00cm, bottom=3.00cm]{geometry}
\usepackage{amssymb}
\usepackage{hyperref}
\hypersetup{                                                          
	bookmarks     = true,            
	pdftoolbar    = true,            
	pdfmenubar    = true,            
	hyperindex    = true,                                                          
	colorlinks    = true,                                                          
	linkcolor     = blue,                                                          
	citecolor     = red,                                                       
	linktocpage   = true,  
	bookmarks     = true,
	linktocpage   = true,
	urlcolor      = red,   
	backref       = true,
}

\newcounter{bla}

\journal{Computer Physics Communications}

\newcommand{\sartre}{Sar{\it t}re }
\newcommand{\dint}{{\rm d}}
\newcommand{\xpom}{x_{I\!\!P}}
\usepackage{svg}
\begin{document}

\begin{frontmatter}
	
	
	
	\title{Predicting the Exclusive Diffractive Electron-Ion Cross Section at small $x$ with Machine Learning in \sartre}
	
	
	\author[]{Jaswant Singh\corref{author}}
	\author[]{Tobias Toll\corref{author2}}
	
	\cortext[author] {\textit{jaswant.singh@physics.iitd.ac.in} }
	\cortext[author2] {\textit{tobiastoll@iitd.ac.in} }
	\address{Department of Physics, Indian Institute of Technology Delhi, India}
	
	\begin{abstract}
The event generator \sartre has been used extensively for simulations of electron-ion collisions in preparation for the Electron-Ion Collider (EIC). \sartre simulates exclusive diffraction in $e$A collisions, in principle for any nuclear species and exclusive final state, usually a vector meson. The coherent and incoherent cross sections for each process are calculated in the colour dipole model for small $x$ from the first and second moments of the respective amplitude, averaged over initial state spatial configurations. Taking these averages is a very CPU demanding task. In order to function as an efficient event generator, these amplitude moments are saved into lookup tables which are used as input for the event generation, making the latter a very fast process. However, there are many recent and ongoing developments of the dipole models underlying the calculations, both in terms of fits of the model parameters to new data as well as new parametrisations of the dipole or proton geometries. Therefore, it is desirable to have a more flexible method for producing the lookup tables. Here, we propose a method using neural networks which can reduce the table production time by 90\% while retaining the same precision in the resulting cross sections. 
	\end{abstract}
	
	
\end{frontmatter}

\section{Introduction}
\sartre \cite{Toll:2013gda,Toll:2012mb} is a Monte Carlo event generator for exclusive diffraction at small gluon momenta $x$ in $ep$ and $e$A interactions, as well as in ultra-peripheral $pp$, $p$A, and AA interactions. It has been used extensively in preparation for the Electron-Ion Collider (EIC) \cite{Accardi:2012qut, Aschenauer:2014cki, AbdulKhalek:2021gbh, ATHENA:2022hxb, Bernauer:2022moy}, as well as for ultra-peripheral $pp$, $p$A, and AA events at the LHC and RHIC experiments \cite{Sambasivam:2019gdd, Toll:2021tcx}. 
\sartre calculates cross sections using the colour dipole model framework \cite{GolecBiernat:1998js, GolecBiernat:1999qd, Kowalski:2003hm, Kowalski:2006hc, Rezaeian:2012ji, Mantysaari:2018nng, Sambasivam:2019gdd} in which the virtual photon splits up into a quark-antiquark colour dipole whose lifetime far exceeds the interaction time. In order to fully take into account the hadronic substructure of heavy nuclei and hadrons, both the first and the second moment of the interaction amplitude are calculated, averaged over initial state spatial gluon configurations. This is a CPU intense task, as the amplitude moments have to be averaged over hundreds of configurations for the sum to converge. \sartre is therefore divided into two parts: the amplitude moment calculations and the event generation. The results from the demanding calculations of the first and second amplitude moments are stored in lookup tables, which are then used as input for the event generator. 
This method has resulted in a very fast and efficient Monte Carlo event generator for the user. 
However, the last years have seen much development in the underlying colour dipole models. New inclusive DIS data released from the H1 and ZEUS experiments at HERA \cite{H1:2015ubc} have resulted in improved fits of the dipole parameters \cite{Mantysaari:2018nng}, as well as a new parametrization suppressing unphysically large dipoles \cite{Sambasivam:2019gdd}. Further developments have included substructure to the hadrons using the so-called hotspot models \cite{Mantysaari:2016ykx, Mantysaari:2016jaz, Kumar:2021zbn}. 

It would be desirable to include these developments of the underlying model and our understanding of the hadronic spatial structure in the \sartre event generator. So far, the stumbling block for the flexibility of the generator lies in the time consuming process of producing the lookup tables for the first two moments of the amplitude. In this work, we propose a method, using machine learning, that will improve the production time of lookup tables by an order of magnitude. We will verify the new method, by making comparisons to existing lookup tables for gold (A=197) and calcium (A=40) targets for several exclusive final states, containing the vector mesons $J/\psi$, $\phi$, or $\rho$. We will thus show the results of six different neural networks. The diversity of these tests gives confidence that the method we propose here can be used for a wide variety of nuclear targets and exclusive final states in \sartre.

The paper is organised as follows. In the next section we describe the theoretical framework used in the calculations of the amplitude moments we wish to store in lookup tables in \sartre. In section \ref{sec:NN} we will describe the neural network model that we are deploying in this work, as well as the data that we fit the network to. In section \ref{sec:results} we show the results of our method, mainly by analysing the resulting relative error in different parts of phase-space. In section \ref{sec:conclusion} we summarise and give a perspective for the future.

\section{Exclusive diffraction in the bSat Dipole Model}
\begin{figure}
	\centering
	\includegraphics[width=0.5\linewidth]{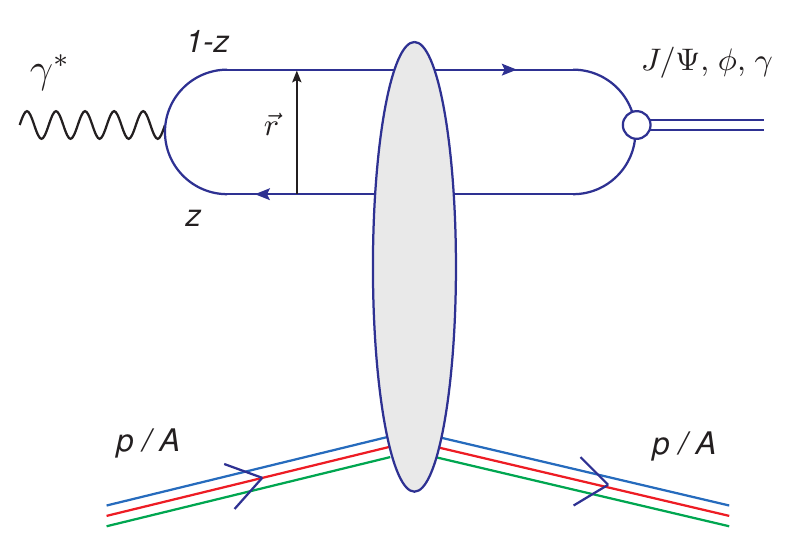}
	\caption{Exclusive vector meson production in the dipole framework of DIS. Description of all variables are given in the text. Figure is from \cite{Kumar:2021zbn}.}
	\label{fig:exclusivedipolemodel}
\end{figure}

In the dipole model, the scattering amplitude for exclusive diffraction in electron-proton scattering factorises into three subprocesses at high energies, as depicted in Fig.\ref{fig:exclusivedipolemodel}. First, the virtual photon splits into a quark-antiquark colour dipole, which then interacts via one or many two-gluon exchanges with the hadronic target. Lastly, the colour dipole recombines into a final state vector meson (or real photon in DVCS). The amplitude is given by \cite{Kowalski:2006hc}:
\begin{eqnarray}
	\mathcal{A}_{T,L}^{\gamma^* p \rightarrow J/\Psi p} (x_{I\!\!P},Q^2,\vb{\Delta})=i\int \dint^2 \vb{r}\int \dint^2\vb{b}\int \frac{\dint z}{4 \pi} (\Psi^*\Psi_V)_{T,L}(Q^2,\vb{r},z) e^{-i[\vb{b}-(1-z)\vb{r}].\vb{\Delta}} \frac{\dint\sigma _{q\bar{q}}}{\dint^2\vb{b}}(\textbf{b},\vb{r},x_{I\!\!P})
	\label{eq:amp}
\end{eqnarray}

Here, $T$ and $L$ represent the transverse and longitudinal polarisations of the virtual photon which carries virtuality $Q^2$, {\bf r} is the transverse size and direction of the dipole, $z$ the energy fraction of the photon taken by the quark, {\bf b} is the impact parameter of the dipole relative the centre of the proton, and $\vb\Delta$ is the exchanged transverse momentum between the proton and the dipole, which is related to the Mandelstam $t$ variable as $|\vb{\Delta}|=\sqrt{-t}$. The overlap between the wave functions of the incoming photon and the outgoing vector meson is denoted $ (\Psi^*\Psi_V)_{T,L}$. The dipole cross section $\dint\sigma/\dint^2\vb{b}$ contains all the information about the target. This amplitude is a Fourier transform from coordinate space to momentum space $\vb\Delta$.

The differential cross section is given by:
\begin{eqnarray}
	\frac{\dint \sigma}{\dint t}^{\gamma^* p \rightarrow J/\Psi p} = \frac{1}{16 \pi} \left<\big| \mathcal{A}^{\gamma^* p \rightarrow J/\Psi p}\big|^2 \right>_\Omega 
\end{eqnarray}
which is the second moment of the amplitude with respect to the initial gluon configuration of the target, here denoted $\Omega$.
This may be expanded in terms of first and second moments as:
\begin{eqnarray}
	\frac{\dint \sigma}{\dint t}^{\gamma^* p \rightarrow J/\Psi p} &=& 
	\frac{1}{16 \pi}  \left|\left<\mathcal{A}^{\gamma^* p \rightarrow J/\Psi p} \right>_\Omega \right|^2 +
	\frac{1}{16 \pi} \left[\left<\big| \mathcal{A}^{\gamma^* p \rightarrow J/\Psi p}\big|^2 \right>_\Omega -
	\left|\left<\mathcal{A}^{\gamma^* p \rightarrow J/\Psi p} \right>_\Omega \right|^2    \right] \nonumber\\
	&=& \frac{\dint \sigma_{\rm coh}}{\dint t}^{\gamma^* p \rightarrow J/\Psi p}+\frac{\dint \sigma_{\rm inc}}{\dint t}^{\gamma^* p \rightarrow J/\Psi p}
\end{eqnarray}
where in the last step we have used the Good-Walker picture \cite{Good:1960ba} in which the first term on the right hand side is identified as the coherent cross section while the term in the square brackets is identified as the incoherent cross section, which is the variance of the amplitude with respect to the initial state.

In the bSat dipole model (also known as IP-Sat) \cite{Kowalski:2006hc}, the dipole cross section is given by:
\begin{eqnarray}
	\frac{\dint \sigma^{(p)}_{q\bar{q}}}{\dint^2\textbf{b}}(\textbf{b},\textbf{r},x_{I\!\!P})=
	2\big[1-\text{exp}\big(-F(x_{I\!\!P},\textbf{r}^2)T_p(\textbf{b})\big)\big],
	\label{eq:bsat}
\end{eqnarray}
where
\begin{eqnarray}
	F(x_{I\!\!P} ,\textbf{r}^2)=\frac{\pi^2}{2N_C} \textbf{r}^2 \alpha_s(\mu^2) x_\mathbb{P} g(x_{I\!\!P},\mu^2).
\end{eqnarray}
This cross section saturates for large $r$ and gluon densities $g$. 
The scale at which the strong coupling $\alpha_s$ and gluon density $g$ are evaluated is $\mu^2 = \mu_0^2 +\frac{C}{r^2}$ and the gluon density at the initial scale $\mu_0$ is parametrised as:
\begin{eqnarray*}
	x g(x,\mu_0^2)= A_g x^{-\lambda_g}(1-x)^{6}
\end{eqnarray*}
where the parameters $A_g, \lambda_g, C$, and $m_f$ are determined through fits to inclusive reduced cross section measurements. 
The bNonSat model is a linearised version of the bSat model with
\begin{equation}
	\frac{\dint \sigma^{(p)}_{q\bar{q}}}{\dint^2\textbf{b}}(\textbf{b},\textbf{r},x_{I\!\!P})=\frac{\pi^2}{N_C}\textbf{r}^2\alpha_s(\mu^2) x_{I\!\!P} g(x_{I\!\!P},\mu^2)  T_p(\textbf{b})
\end{equation}
which does not saturate for large gluon densities and large dipoles. This dipole cross section corresponds to a single two-gluon exchange. 
The transverse profile of the proton is usually taken to be Gaussian:
\begin{eqnarray}
	T_p(\textbf{b}) = \frac{1}{2 \pi B_G}\exp\bigg(-\frac{\textbf{b}^2}{2B_G}\bigg)
	\label{eq:profile}
\end{eqnarray}
This represents a proton without internal structure. In this case, the variance of the amplitude becomes zero as the first (squared) and second moments of the amplitude coinside. However, the proton can be given substructure by modifying the thickness function, e.g. in the hot-spot models \cite{Mantysaari:2016ykx, Mantysaari:2016jaz, Kumar:2021zbn}:
\begin{eqnarray}
	T_p(\textbf{b}) = \frac{1}{N_{q}}\sum_{i=1}^{N_{q}}T_{q}(\textbf{b}-\textbf{b}_i),
\end{eqnarray}
where the $N_q$ hotspots each have a Gaussian profile, and their positions $\vb{b}_i$ are sampled from a Gaussian such that the average profile coinside with \eqref{eq:profile}. This procedure enables the calculation of incoherent cross sections in $ep$ scattering.

\subsection{The Dipole Model in $e$A}
For extending the dipole model to heavy nuclei, one may adopt the independent scattering approximation:
\begin{eqnarray}
	1-\mathcal{N}^{\rm (A)}(\xpom, \vb{r}, \vb{b})=\prod_{i=1}^A\left(1-\mathcal{N}^{\rm (p)}(\xpom, \vb{r}, |\vb{b}-\vb{b}_i|)\right)
	\label{eq:indscat}
\end{eqnarray}
where the scattering amplitude $\mathcal{N}=\frac12 \frac{ \dint \sigma _{q\bar{q}}}{\dint^2\vb{b}}$. Here, the positions $\vb{b}_i$ are sampled from a Woods-Saxon distribution. Plugging eq.\eqref{eq:bsat} into eq.\eqref{eq:indscat} gives:
\begin{eqnarray}
	\frac{\dint \sigma^{\rm (A)}_{q\bar{q}}}{\dint^2\textbf{b}}(\textbf{b},\textbf{r},x_{I\!\!P})=
	2\big[1-\text{exp}\big(-F(x_{I\!\!P},\textbf{r}^2)\sum_{i=1}^nT_p(|\vb{b}-\vb{b}_i|)\big)\big],
\end{eqnarray}
This may be seen as a modification of the thickness function for heavy ions, being expressed as the sum of the thickness functions of all the nucleons. 

\subsection{The Dipole Model in \sartre}
In the event generator \sartre \cite{Toll:2013gda}, the coherent and incoherent cross sections are calculated by taking the average of nucleon (or nucleon and hotspot) configurations for an observable $\mathcal{O}$ as:
\begin{eqnarray}
	\left<\mathcal{O}\right>_\Omega = \frac{1}{N}\sum_{i=1}^N \mathcal{O}(\Omega_i)
	\label{eq:average}
\end{eqnarray}
where $\Omega_i$ represents a single gluon-configuration in transverse space. 
When considering the $t$-spectrum of the differential cross sections one may notice that at small $|t|$, the cross section is completely dominated by the coherent interactions, while already at moderate $|t|\sim 1/R_A^2$, with $R_A$ being the radius of the nucleus, the incoherent cross section begins to dominate. This means that the sum in eq.\eqref{eq:average} for the first moment of the amplitude converges very fast for small $|t|$, while for larger $|t|$ the variance of the amplitude becomes much larger than its average, and the convergence of the sum becomes rather poor. For the second moment, the situation is better for larger $|t|$ and it was found that good convergence for $|t|<0.5~{\rm GeV}^2$ is achieved with $N=500$ \cite{Toll:2013gda}. 
For the first moment of the amplitude, one may note that it is given by:
\begin{eqnarray}
	\left<\mathcal{A}_{T, L}(Q^2, \Delta, \xpom)\right>_\Omega= 
	\int\pi r\dint r\dint z\b \dint b (\Psi_V^*\Psi)_{T, L}(Q^2,r,z) J_0([1-z]r\Delta/2)
	\left<\frac{\dint \sigma_{q\bar q}}{\dint^2\vb{b}}\right>_\Omega(\xpom, r, b)
\end{eqnarray}
and one may write to a good approximation \cite{Kowalski:2003hm}:
\begin{eqnarray}
	\left<\frac{\dint \sigma_{q\bar q}}{\dint^2\vb{b}}\right>_\Omega(\xpom, r, b)=
	2\left[ 1-\left(
	1-\frac{T_{\rm WS}(\vb{b})}{2}\sigma^{(p)}_{q\bar q}(\xpom, r)
	\right)^A\right]
\end{eqnarray}
where $T_{\rm WS}$ is the two-dimensional Woods-Saxon profile. Here
\begin{eqnarray}
	\sigma^{(p)}_{q\bar q}(\xpom, r)=\int\dint^2\vb{b}\frac{\dint\sigma^{(p)}}{\dint^2\vb{b}}(\xpom, \vb{r}, \vb{b})=
	4\pi B_G^2 (\ln(G)-Ei(-G)+\gamma_{\rm Euler})
\end{eqnarray}
with $G=(\pi r^2\alpha_S(\mu^2)\xpom g(\xpom, \mu^2))/(4N_C B_G)$.

\section{Fitting the second moment of the $e$A amplitude with a Neural Network}
\label{sec:NN}
\sartre generates events by generating the kinematic variables $Q^2$, $W^2$, and $t$ from the total cross section. Here, $Q^2$ is the virtuality of the virtual photon and $t$ is the Mandelstam variable. $W$ is the invariant mass of the photon-proton/nucleus interaction given by:
\begin{eqnarray}
W^2=Q^2\frac{1-\xpom}{\xpom}+\frac{M_V^2}{\xpom}+m^2_N,
\end{eqnarray}
where $M_V$ is the mass of the vector meson and $m_N$ the nucleon mass. In order to calculate the second moment of the cross section in a point in phase-space $(Q^2, W^2, t)$
one needs to numerically evaluate the five-dimensional integral in eq.\eqref{eq:amp} one thousand times (500 for each photon polarisation).
For an event-generator this procedure becomes prohibitively slow. In order to create a usable event generator, lookup tables for the first and second moment of the amplitude, as well as their difference, are created separately and saved to disk. This is a process that can take many months on computer farms. For generating events one then interpolates for any kinematic point in $(Q^2, W^2, t)$ from the lookup tables and generate events with the correct distributions at a very fast rate. This process has to be repeated for each initial state (nuclear species) and for each final state ($\Upsilon$, $J/\psi$, $\phi$, $\rho$, etc.). Further, the dipole model is an area of active research, with new versions and fits being published regularly. Every time a new process or version of the model is to be implemented, new tables need to be created, which is a demanding task taking many person- and CPU-hours to complete. 

The lookup tables in \sartre are stored in 3D histograms in ROOT format (TH3F)\cite{Brun:1997pa}.
When \sartre is generating events using the lookup tables as input, the event generator interpolates between the bins nearest to the kinematic point. However, this poses a problem, as there is no clearly defined metric on the lookup table. Even though the three dimensions of the tables ($Q^2$, $W^2$, and $t$) all carry dimension GeV$^2$, their physical meanings and ranges are vastly different. The interpolation procedure relying on nearest bins is therefore ill defined.

In the following we introduce an alternative way to generate amplitude lookup tables as well as events using machine learning with neural networks. We will demonstrate that this method reduces the production time of new tables by an order of magnitude. This also enables a more consistent interpolation procedure as the neural networks are using the entire table when it performs an interpolation.

The proposed method has three steps. First, one generates a small lookup table for the second moment of the dipole amplitude using the method described above. Then one fits a neural network to the small table. This is then a parametrisation of the bSat dipole model. In the last step one uses the neural network to create a new table with fine enough binning to generate events in the desired phase-space. One may also, as an alternative, use the resulting neural network as an input for \sartre directly instead of using lookup tables. This solution would provide a more consistent method of interpolation between data points, but would not be backwards compatible with previous versions of \sartre\!\!.

In the following we will demonstrate the efficacy of this method, and find its limits. We develop the method using lookup tables for $e+$Au$\rightarrow e'+$Au$'+J/\psi$. We will then apply the developed method, as a cross check, to more exclusive final states, namely $\phi$ and $\rho$ mesons, as well as for calcium targets.


\subsection{Model and data description}
As a training dataset we use the existing tables in \sartre using the parametrisation in \cite{Kowalski:2006hc} for the bSat model. These tables come in two regions of $t$, where the $t$-binning is finer for smaller $|t|$. We use 12 sets of tables, with $\gamma^*_P+N\rightarrow V+N'$, where $P$ is longitudinal or transverse polarisation, the nuclear species $N$ is either gold or calcium, and the final state vector meson $V=J/\psi$, $\phi$, or $\rho$. These data tables are summarised in tables \ref{tab:Au} and \ref{tab:Ca} for gold and calcium targets respectively.
\begin{table}
\begin{center}
	\begin{tabular}{ |c| c| c| c | c | c |}
		\hline
 		Variable & Min. & Max. & \#bins & Scale & Region  \\\hline
 		$Q^2 $ & $10^{-4}$ & $20$ & 20 &Logarithmic & 1 and 2 \\ \hline
		$W^2$ & $M_V^2$ & 20164 & 55 & Linear & 1 and 2 \\ \hline
		$t$ & -0.051 & 0 & 17 & Linear & 1 \\ \hline
		$t$ & -0.5 & -0.03 & 20 & Linear & 2 \\\hline
	\end{tabular}
	\caption{A description of the the data tables of the second moment of the amplitudes for $\gamma^*+{\rm Au}\rightarrow V+{\rm Au}$, with $V=J/\psi$, $\phi$, or $\rho$. Here, $M_{J/\psi}^2=10.4712~{\rm GeV}^2$, $M_{\phi}^2=1.91964~{\rm GeV}^2$, and $M_{\rho}^2=1.48174~{\rm GeV}^2$. All units are in GeV$^2$.}
	\label{tab:Au}
\end{center}
\end{table}

\begin{table}
\begin{center}
	\begin{tabular}{ |c| c| c| c | c | c |}
		\hline
 		Variable & Min. & Max. & \#bins & Scale & Region  \\\hline
 		$Q^2$ & $10^{-4}$ & $20$ & 20 &Logarithmic & 1 and 2 \\ \hline
		$W^2$ & $M_V^2$ & 20164 & 55 & Linear & 1 and 2 \\ \hline
		$t$ & -0.06 & 0 & 10 & Linear & 1 \\ \hline
		$t$ & -0.8 & -0.06 & 20 & Linear & 2 \\\hline
	\end{tabular}
	\caption{A description of the the data tables of the second moment of the amplitudes for $\gamma^*+{\rm Ca}\rightarrow V+{\rm Ca}$, with $V=J/\psi$, $\phi$, or $\rho$. Here, $M_{J/\psi}^2=10.4712~{\rm GeV}^2$, $M_{\phi}^2=1.91964~{\rm GeV}^2$, and $M_{\rho}^2=1.48174~{\rm GeV}^2$. All units are in GeV$^2$.}
	\label{tab:Ca}
\end{center}
\end{table}

In any region of overlap between the tables, the table with finer binning is prioritised. We train the neural networks to both regions simultaneously. 
Before training we normalise the table data such that all variables are in the range $[0,1]$:
\begin{eqnarray}
	X_{\rm scaled} =  \frac{X_{\rm old}-X_{\rm min}}{X_{\rm max}-X_{\rm min}} 
\end{eqnarray}
where $X$ is the variable being scaled, and $X_{\rm old}$ represent the grid points of the tables.

We use a subset of the bins in the table to train the network, ranging from 1\% of the bins upto 20\%. Out of these, we use 80\% of the bins for training the network and 20\% for validation, and the Early Stopping method for avoiding overfitting of the data. We found a good convergence for a neural network architecture with three input nodes for the kinematic point in $Q^2$, $W^2$, and $t$, and one output node for the second moment of the amplitudes $\left<\mathcal{A}_{L, T}^2\right>$, with 3 hidden layers inbetween, each with 30 nodes. 
The output of node $j$ in layer $i$ is given by the activation function as $x^i_j=\tanh(z^i_j)$ where the weighted sum $z^i_j$ is defined as:
\begin{eqnarray}
	z^i_j=\sum_k w^i_{jk}x^{i-1}_k+b^i_j.
\end{eqnarray}
Here, $w^i_{jk}$ denotes the weight of the link between node $j$ in layer $i$ and node $k$ in layer $i-1$, and each node has a bias $b^i_j$. The weights and biases are initiated with random values and we use the Back-Propagation \cite{Rumelhart:1986} algorithm to fit them to the data. 

\section{Results}
\label{sec:results}
\begin{figure}
	\centering
	\includegraphics[width=0.5\linewidth]{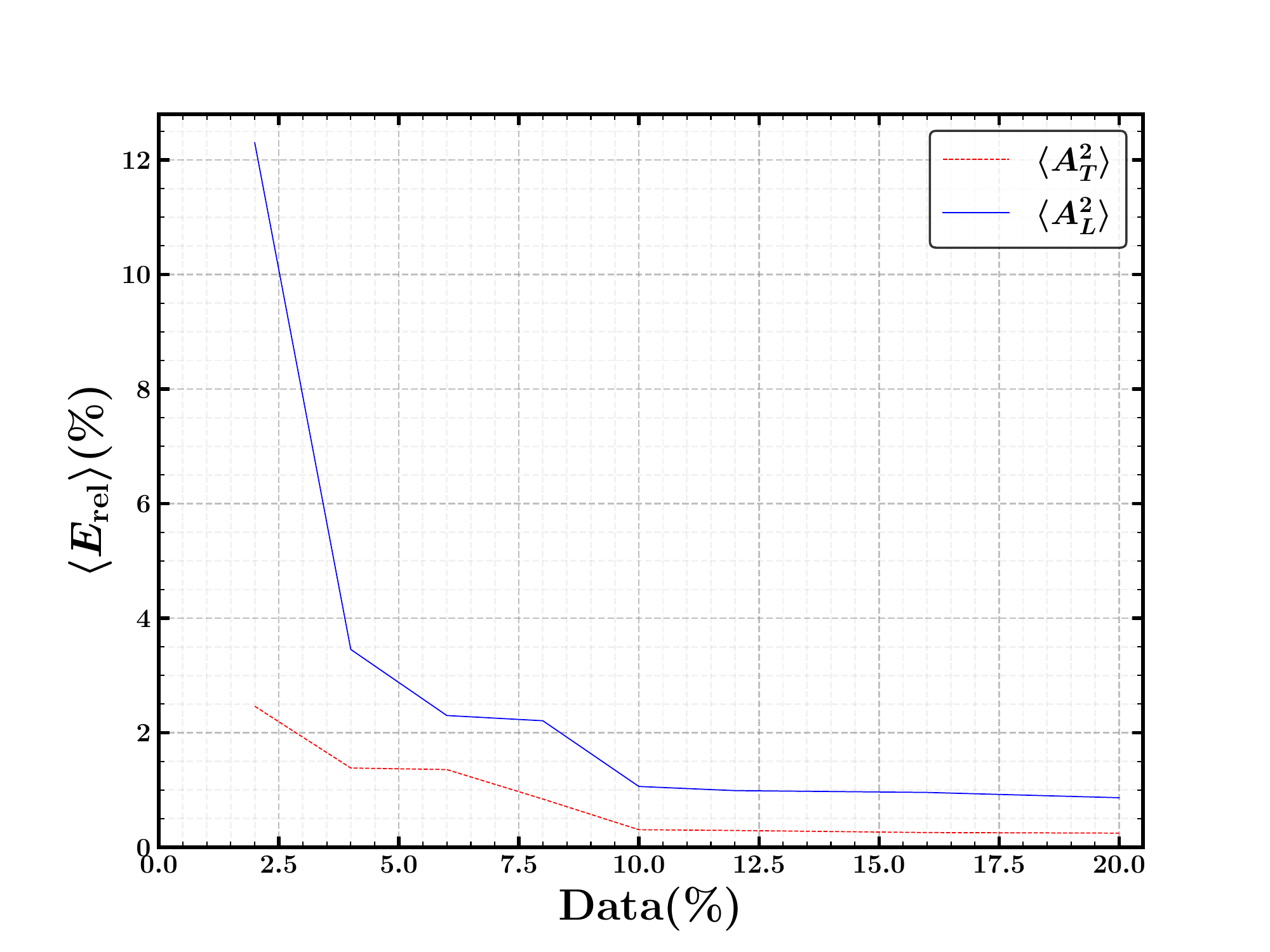}
	\caption{Average relative error (in \%) of the neural network's prediction of the entire table as a function of the percentage of the table it has been trained on.}
	\label{fig:relativeerrora197}
\end{figure}

In Fig.~\ref{fig:relativeerrora197} we show the results from our trained neural network on the data set described above. 
The plots show the average relative error of the resulting network in describing the entire table, including the points which were not part of the training, against the percentage of table bins included in the training. The average relative error is defined by:
\begin{eqnarray}
	\left<E_{\rm rel}\right>=\frac{1}{N}\sum_i^N\frac{
		\left| \left<\mathcal{A}_{L,T}^2\right>_{\Omega}^{\rm (NN)} -  \left<\mathcal{A}_{L,T}^2\right>_{\Omega}^{\rm (table)}
		\right|}{ \left<\mathcal{A}_{L,T}^2\right>_{\Omega}^{\rm (table)}}
		\label{eq:relerr}
\end{eqnarray}
where $N$ is the total number of data points in the table. We see that the relative error in predicting the entire table drops quickly as we include more points in the training, and we get a small relative error already when we train on 10\% of the data. This means that we potentially can reduce the production time for table creation by 90\%. We also see that the relative error is in general larger for the longitudinally polarised photons than for the transverse polarisation. 

\begin{figure}
	\centering
	\includegraphics[width=0.95\linewidth]{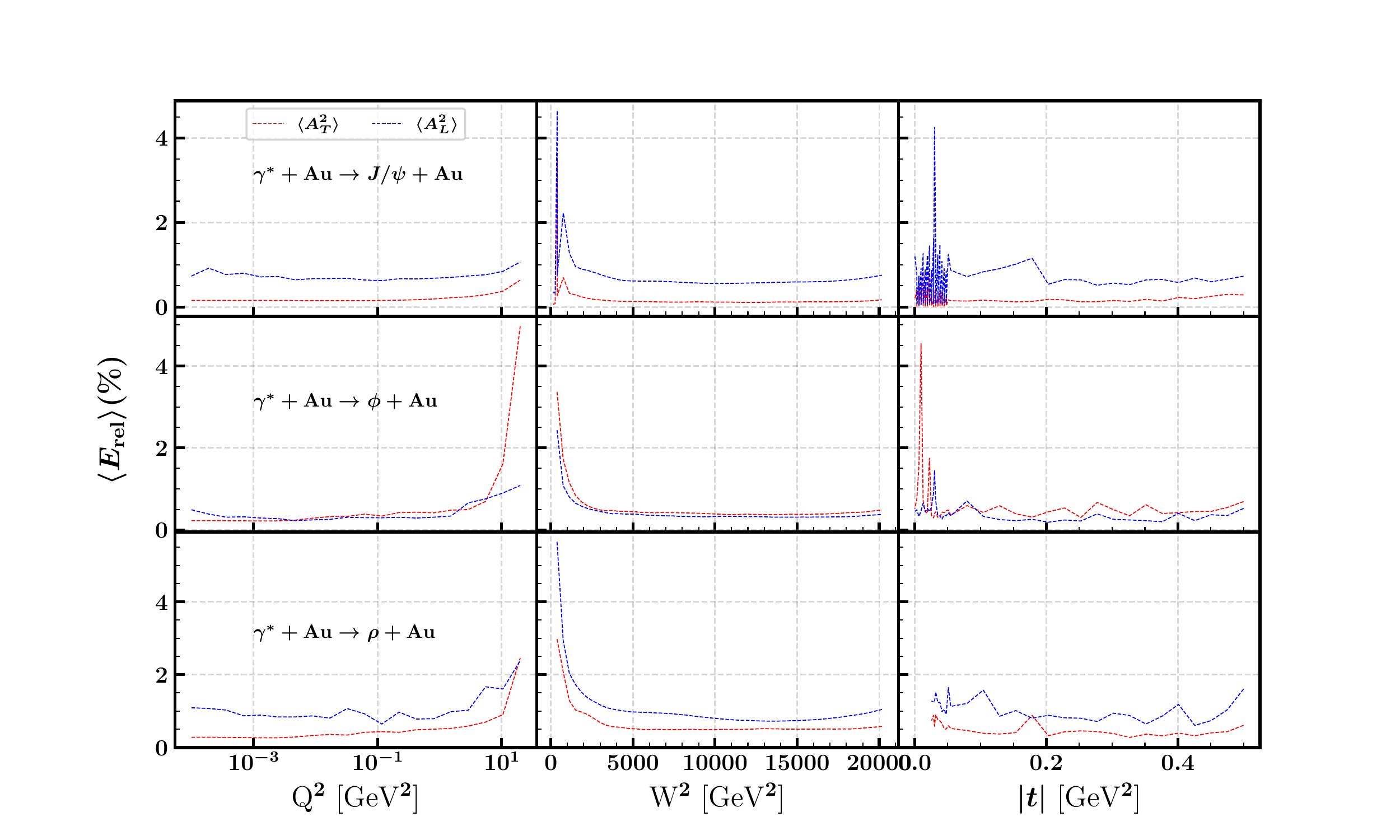}
	\caption{Average relative error (in \%) as a function of $W^2$, $Q^2$, and $t$ of the neural networks' prediction of the entire table. Each row shows the result of a neural network that has trained on prediction for $J/\psi$, $\phi$, and $\rho$ production respectively. Each network has been trained on 10\% of the bins in the original respective tables.}
	\label{fig:Au-relerr}
\end{figure}

In figure \ref{fig:Au-relerr},  we show the relative error of the neural network which has been trained on 10\% of the data points in predicting $Q^2$, $W^2$, and $|t|$ respectively. We do this for all three vector meson species considered in this work, namely $J/\psi$, $\phi$, and $\rho$ mesons. Here, we fix the bin in one of the kinematic variables and have the sum in eq.\eqref{eq:relerr} run over the bins in the other two variables. We see that the relative error is flat in all three distributions for all three final states, except where the cross section becomes small, in which case there is a minor spike in the relative error. The square amplitudes for longitudinally polarised photons are consistently larger than for transverse polarisation. This is likely due to the $Q^2$ dependence being more complex in the former case. It should also be noted that the longitudinal amplitude's second moment is smaller than in the transverse case. 

\begin{figure}
	\centering
	\includegraphics[width=0.95\linewidth]{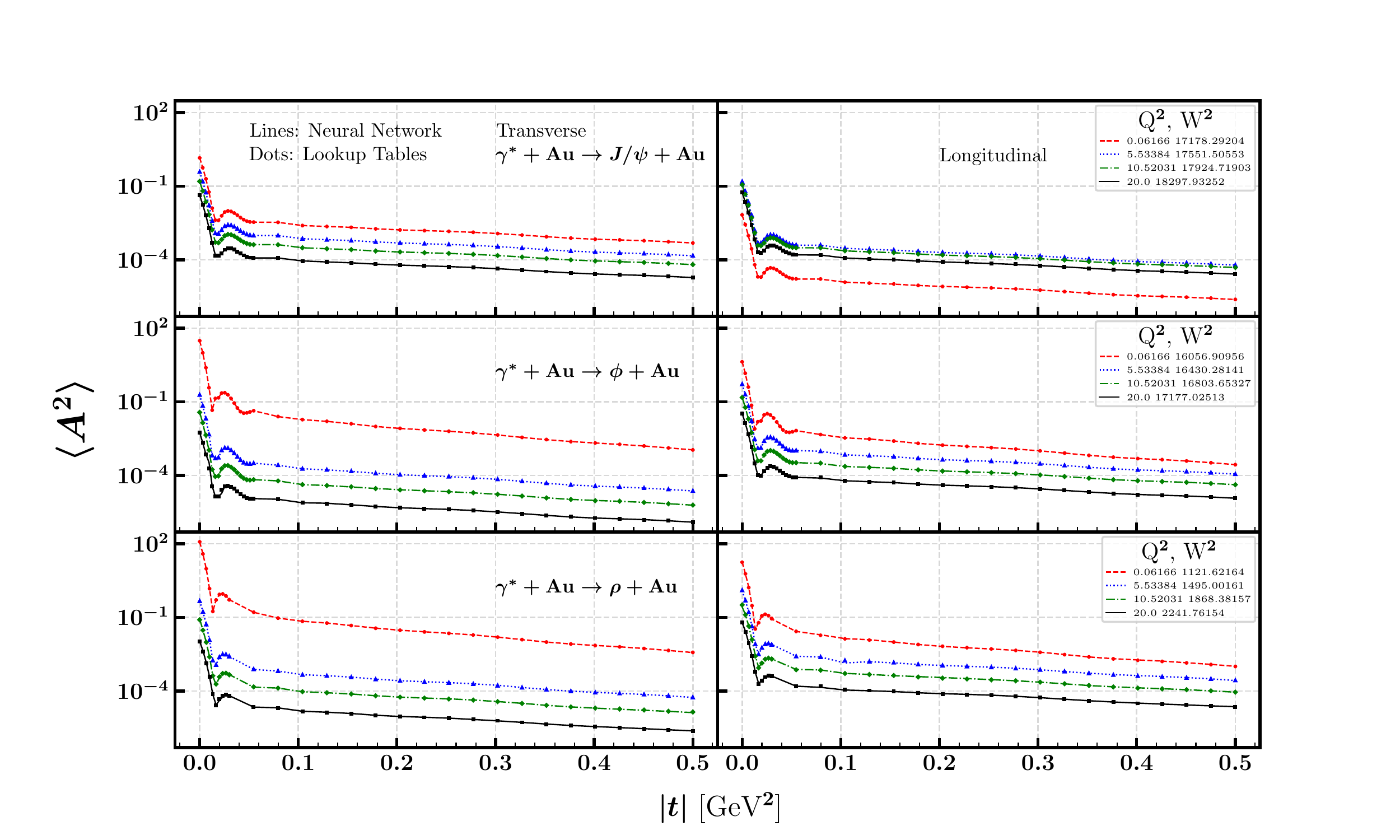}
	\caption{The $t$-spectra of the second moments of the amplitude for a few values of $Q^2$ and $W^2$. Left panels show the results for transverse polarised photons, and right panels for longitudinal polarisation. The top row show the results for $J/\psi$ production, middle for $\phi$ production, and bottom for $\rho$ production. Neural networks are trained on 10\% of the table data.}
	\label{fig:Au-t}
\end{figure}

In figure \ref{fig:Au-t} we show the resulting $t$-spectra for four combinations of $Q^2$ and $W^2$, and compare them with the original table results. Here, the neural networks are trained on 10\% of the table data and is describing the $t$-spectrum well, showing that we can have confidence in our method to accurately generate events. 

\begin{figure}
	\centering
	\includegraphics[width=0.95\linewidth]{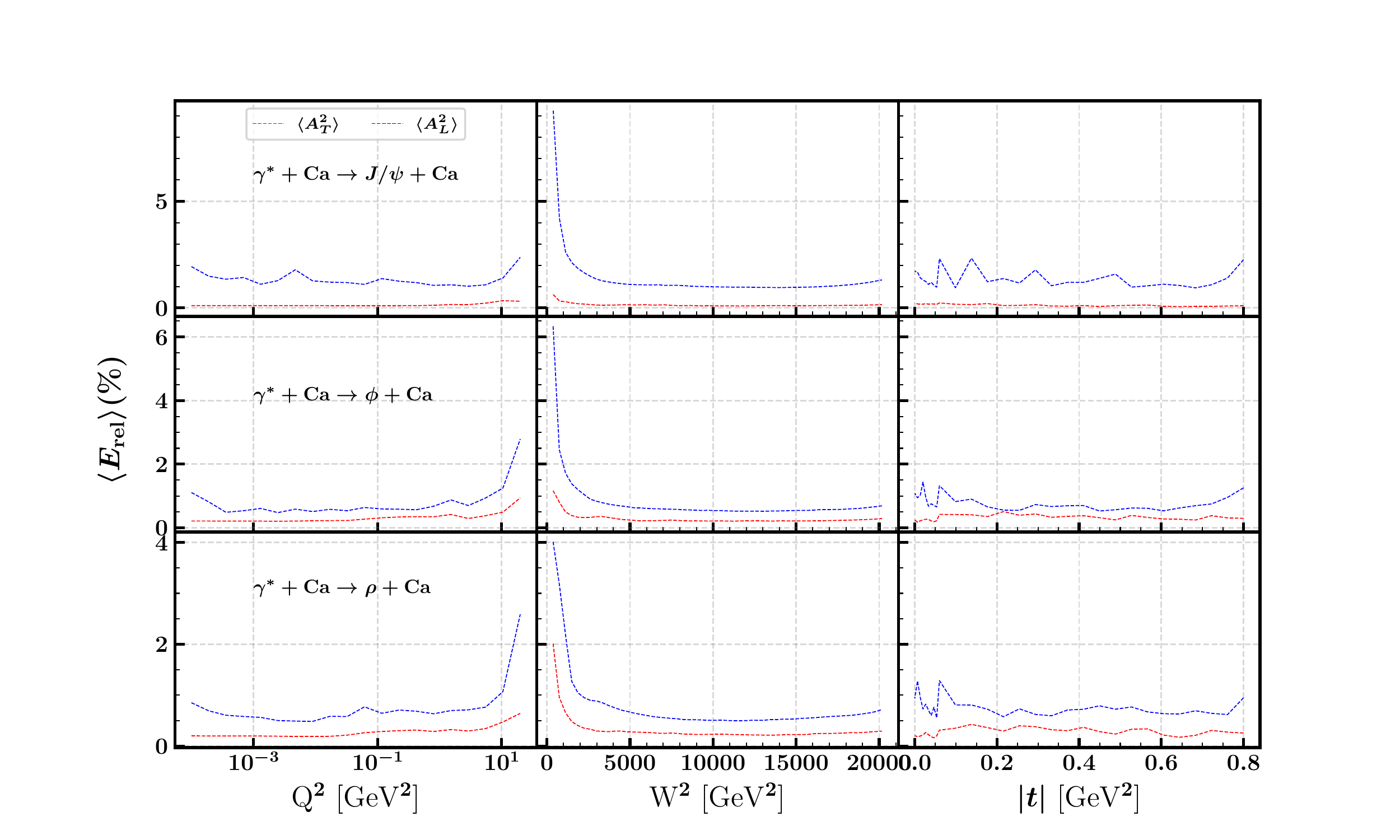}
	\caption{Average relative error (in \%) as a function of $W^2$, $Q^2$, and $t$ of the neural networks' prediction of the entire table. Each row show the result of a neural network that has trained on predictions for $J/\psi$, $\phi$, and $\rho$ production respectively. Each network has been trained on 10\% of the bins in the original respective tables.}
	\label{fig:Ca-relerr}
\end{figure}

\begin{figure}
	\centering
	\includegraphics[width=0.95\linewidth]{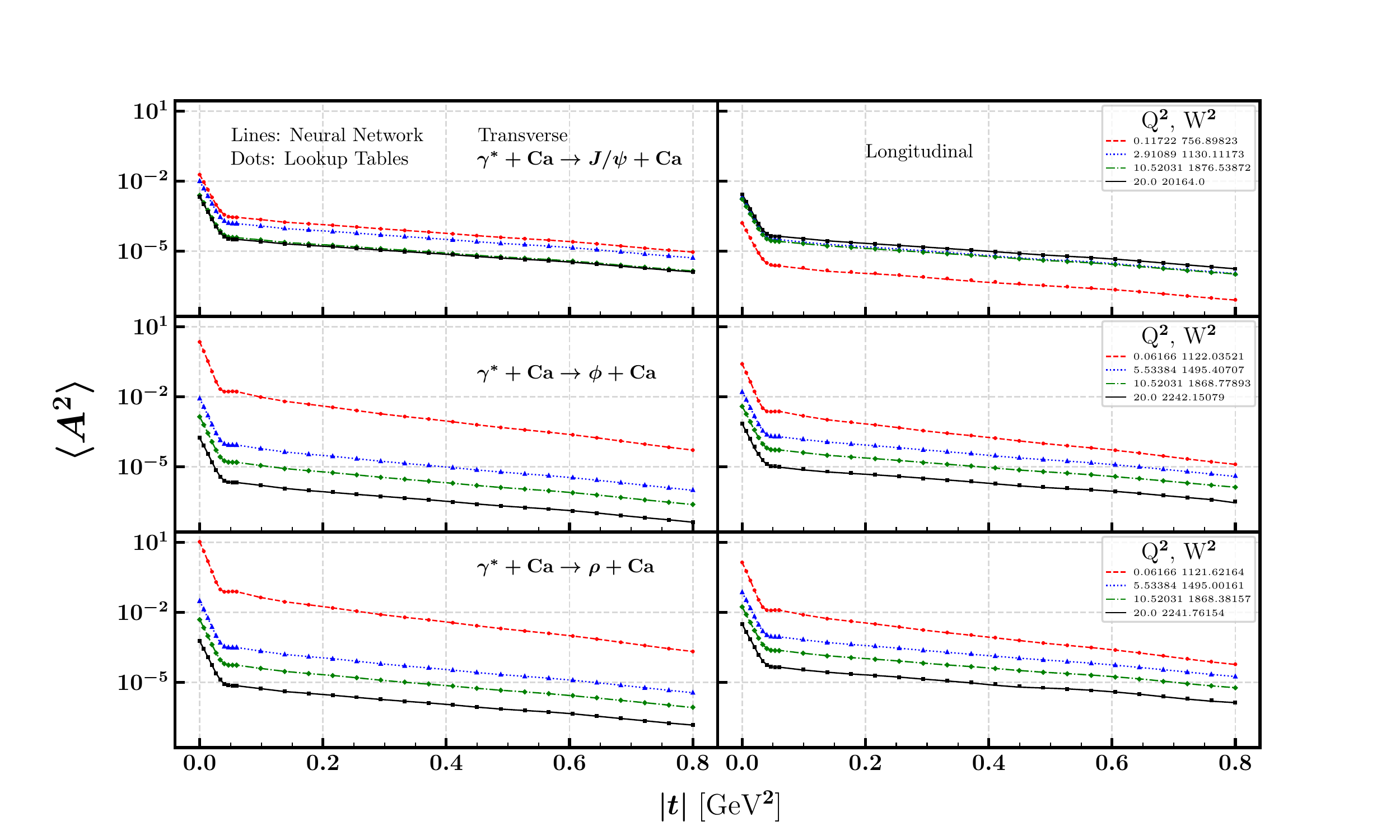}
	\caption{The $t$-spectra of the second moments of the amplitude for a few values of $Q^2$ and $W^2$. Left panels show the results for transverse polarised photons, and right panels for longitudinal polarisation. The top row show the results for $J/\psi$ production, middle for $\phi$ production, and bottom for $\rho$ production. Neural networks are trained on 10\% of the table data.}
	\label{fig:Ca-t}
\end{figure}

In figs. \ref{fig:Ca-relerr} and \ref{fig:Ca-t}, we have applied the same method for calcium targets. We have again trained the neural networks on 10\% of the lookup table data, and we see that the performance of the resulting neural networks is consistent with that for gold targets. Here as well, the relative errors are below a few percent over the whole spectrum and becomes as large as 6\% in kinematic points where the absolute value of the the amplitude's second moment is the smallest. This gives us confidence that the method described in this paper can be applied to a wide variety of both target nuclear species as well as exclusive final states. 


\section{Conclusions and Outlook}
\label{sec:conclusion}
\sartre is a very fast event generator for exclusive diffractive events at small $x$, using the colour dipole model. However, a practical limitation of the generator is that it relies on lookup tables for the first and second moment of the amplitude. These tables are separate for longitudinal and transversely polarised photons, for each nuclear species, and for each final state vector meson. These tables may take months of CPU and people power intense work on computer farms to produce. In this work, we have demonstrated that the production time for these tables can be reduced by an order of magnitude if we fit neural networks to the model. In the proposed method, one produces  small tables, containing only 10\% of the data-points using the previous method. We then fit neural networks to the produced tables, and use the resulting networks to fill in the remaining 90\% of the data points. We have demonstrated that this method give small relative errors compared to the default method, around or below 1\% in most of phase-space and only reaching around 5\% in kinematic regions where the vector meson production cross section is small. Furthermore, this method is completely backwards compatible with earlier versions of \sartre, as it only modifies the method by which the tables are produced, and not how the tables are used for event generation. We envision a future version of \sartre which may use neural networks directly as an input for event generation as well. This would solve the problem of ambiguity for interpolating on tables without a well defined metric.

\section{Acknowledgments}
We thank T. Ullrich for helpful comments which improved the paper. The work of JS is supported by the University Grants Commission, India, UGC-Ref. No.:1218/(CSIR-UGC NET DEC 2018). We thank the physics department at IIT Delhi for support.

\bibliographystyle{elsarticle-num}
\bibliography{draft_neural_network_v5.bib}

\end{document}